\documentclass[a4paper, 12pt]{article} 
\usepackage[right=1.1in,left=1.1in,top=1.1in,bottom=1.1in]{geometry}
\usepackage{graphicx} 
\usepackage{braket}
\usepackage{mathrsfs}
\usepackage[T1]{fontenc} 
\usepackage{amsmath}
\usepackage{pxfonts}
\usepackage[T1]{fontenc}
\usepackage{amssymb}
\usepackage{natbib}
\usepackage{epstopdf}
\usepackage{setspace}
\usepackage{enumitem}
\usepackage{sectsty}
\usepackage{wrapfig} 
\sectionfont{\large}
\bibpunct{(}{)}{,}{a}{}{,}
\usepackage{tcolorbox}
    \usepackage{pgfplots}

\newcommand{\Hilbert}{\mathscr{H}}
\newcommand{\tr}{\text{tr}}
\newcommand{\Prob}{\text{Prob}}

\usepackage{tikz}   
\usepackage{tikz-3dplot} 

\newtheorem{theorem}{Theorem}[section]

\newcommand{\qed}{\nobreak \ifvmode \relax \else
      \ifdim\lastskip<1.5em \hskip-\lastskip
      \hskip1.5em plus0em minus0.5em \fi \nobreak
      \vrule height0.75em width0.5em depth0.25em\fi}

\usepackage{bbm}
\usepackage{MnSymbol}
\usepackage[all]{xy}

\usepackage{xcolor,colortbl,array,amssymb}

\usepackage{epigraph}
\makeatletter

\newlength{\bibitemsep}
\newlength{\bibparskip}
\let\oldthebibliography\thebibliography
\renewcommand\thebibliography[1]{%
  \oldthebibliography{#1}%
  \setlength{\parskip}{\bibitemsep}%
  \setlength{\itemsep}{\bibparskip}%
}

\title{\textbf{Density Matrix Realism}\footnote{Penultimate version, chapter in Cuffaro, M. E., and Hartmann, S. (eds.), \textit{The Open Systems View: Physics, Metaphysics and Methodology}, Oxford University Press, forthcoming}} 

\author{Eddy Keming Chen\thanks{Department of Philosophy,  University of California, San Diego, 9500 Gilman Dr, La Jolla, CA 92093-0119. Website: www.eddykemingchen.net. Email: eddykemingchen@ucsd.edu  }}

\date{\today}

\begin{document}
\bibliographystyle{apa}

\maketitle 



\begin{abstract}
Realism about quantum theory naturally leads to realism about the quantum state of the universe. It leaves open whether it is a pure state represented by a wave function, or an impure one represented by a density matrix. I characterize and elaborate on Density Matrix Realism, the thesis that the universal quantum state is objective but \textit{can be impure}. To clarify the thesis, I compare it with Wave Function Realism, explain the conditions under which they are empirically equivalent,  consider two generalizations of Density Matrix Realism, and answer some frequently asked questions. I end by highlighting an implication for scientific realism. 

\end{abstract}

\hspace*{3,6mm}\textit{Keywords: empirical equivalence, local indistinguishability, wave function realism, scientific realism, canonical typicality, quantum entanglement, conditional density matrix, reduced density matrix, collapse, Wentaculus, open systems} 
\begingroup
\singlespacing
\tableofcontents
\endgroup

\section{Introduction}

Realism about quantum theory naturally leads to realism about the quantum state of the universe, the idea that the universal quantum state is an objective and mind-independent feature of the physical world. It leaves open, however, whether it is a pure state represented by a wave function, or an impure (mixed) one represented by a density matrix. In this paper, I characterize and elaborate on Density Matrix Realism (DMR), the thesis that the universal quantum state is objective but \textit{can be impure}. 

 Related ideas have been considered in quantum foundations \citep{durr2005role, maroney2005density, tumulka2010bohmian, allori2013predictions, wallace2012emergent, chen2022uniform}, quantum cosmology \citep{page1986density, page2008no, barvinsky2006cosmological},  quantum gravity \citep{hawking1976breakdown, hawking1982unpredictability}, and philosophy of science \citep{McCoySMS, robertson2022search, cuffaro2021open}. 
 
  DMR was formulated and given its name in \cite{chen2018IPH}\footnote{The formulation of DMR in that paper seems to require the universal density matrix be impure, but I intended to allow the possibility that it is pure.}, with further developments in \cite{chen2018valia,chen2018HU,chen2018NV,chen2020harvard,chen2022strong,chen2023detlef}.   DMR provides a broad (and I think fruitful) framework for thinking about quantum theory. 

For various reasons, DMR remains puzzling to many people. The common issues are: 
\begin{enumerate}
  \item Density matrices are often regarded as second-class citizens in quantum theory, representing  subsystem states (in the form of reduced density matrices) or epistemic states (in the form of statistical density matrices).  Such a perspective is taught in standard textbooks on quantum mechanics and assumed in many research articles in quantum foundations. 
  \item  Density-matrix theories of quantum mechanics are less familiar than  wave-function theories. Although both can be adapted to solve the quantum measurement problem, the density-matrix solutions are not widely known.
  \item It is unclear how to interpret the density matrix as an objective feature of the universe. 
  \item It is unclear whether and how density-matrix theories can reproduce the same empirical predictions as wave-function theories. In other words, it is unclear whether and under what conditions such theories are empirically equivalent. 
  \item It is unclear what theories with a universal density matrix look like from the subsystem level. 
  \item There are multiple versions of DMR; it is sometimes unclear which one is being discussed in a particular context. 
\end{enumerate}
This paper attempts to address such issues. First, I discuss some definitions, examples, and interpretations of density-matrix realist theories. Second, I suggest that wave-function theories and density-matrix theories can be empirically equivalent at the universal level, the subsystem level, and the local level.  Third, I consider two generalizations of density matrix realism. Finally, I answer six frequently asked questions and highlight an implication for scientific realism. 

The goal here is to clarify DMR. There are other issues worth considering, regarding the motivations for and arguments in favor  of DMR. Although they are not the focus here, I  briefly address them in \S7.6.  See the companion paper \cite{chen2023detlef} for more related discussions and references.

\section{Definitions}

I begin by defining Wave Function Realism (WFR), a more familiar and popular thesis about quantum state realism, and a close cousin to DMR. 

Consider realist solutions to the quantum measurement problem, such as Bohmian mechanics, objective collapse theories, and Everettian quantum mechanics. In such theories, the universal quantum state plays an indispensable role in the kinematics and the dynamics, providing reasons to regard it as an objective feature of reality. Given how such theories are often presented, it is widely believed that the objective universal quantum state has to be pure, represented by a wave function. Let us formulate the thesis as follows:

\begin{description}
  \item[Wave Function Realism (WFR)] The quantum state of the universe is objective; it has to be pure. 
\end{description}
This characterization of WFR is broader than that of \cite{AlbertEQM} and \cite{ney2021world}. For them, WFR carries a specific commitment to understanding the universal wave function as a physical field that lives on a vastly high dimensional ``configuration'' space, from which the ordinary 3-dimensional space is emergent. That is, however, not the only way to be a realist about the wave function. For example, the multi-field interpretation, spacetime state realism, and  the nomological interpretation all count as realist interpretations of the wave function, and hence properly fall under the same label \citep{chen2019realism}.

WFR, thus defined, is a standard position for realism about the universal quantum state. We often assume that the universal quantum state, if objective, must be pure.  In textbook presentations, mixed states are often used to represent reduced or statistical states, as expressions of entanglement with other systems or our ignorance of the actual pure state. However, there is no compelling reason why the universe cannot be in a fundamental mixed-state density matrix,  one that does not arise from entanglement or lack of knowledge. In fact, we can formulate Bohmian mechanics, objective collapse theories, and Everettian quantum mechanics with a fundamental density matrix \citep{durr2005role, maroney2005density, allori2013predictions, wallace2012emergent}. I explain them in \S3. 

Let us consider an alternative to WFR:
\begin{description}
  \item[Density Matrix Realism (DMR)] The quantum state of the universe is objective; it can be pure or impure. 
\end{description}
This thesis provides additional theoretical freedom in choosing the quantum state of the universe; it can be pure or impure (a ``mixed state''). The option to use impure states is crucial for certain proposals in quantum foundations and quantum cosmology (\S3). 

Let us clarify some key terms in both theses.  

(i) ``The'': it implies uniqueness. Both theses differ from the proposal that seems to be suggested by \cite{wallace2016probability} that the universe can possess two physical states at the same time: a fundamental pure state and a fundamental mixed state. 

(ii) ``Quantum state of the universe'': both theses are about the quantum state of the universe. It does not logically entail that  subsystem quantum states must be objective or that they must be pure (or impure). We consider a generalization to subsystem quantum states in \S6.

(iii) ``Objective'': it means that the universal quantum state corresponds to an objective feature of reality, which is not merely epistemic (encoding lack of knowledge), or pragmatically useful (merely an instrument for calculations). The meaning of objectivity  is left open-ended, leaving room for different realist interpretations \citep{chen2019realism}. We consider several options in \S4. 

(iv) ``Must be pure'' vs. ``Can be pure or impure'': this is the only difference between the two theses. ``Must'' and ``can'' are  modal concepts. WFR restricts universal quantum states to only pure ones, while DMR allows both pure and impure universal quantum states.  However, the latter is compatible with additional laws of physics (such as the Initial Projection Hypothesis proposed in \cite{chen2018IPH}) that make it physically impossible for the universe to be in a pure state.


\section{Examples}

Let us illustrate WFR and DMR with some examples of physical theories. 

 \textbf{$\Psi$-BM.} Consider the standard formulation of Bohmian mechanics ($\Psi$-BM) that validates WFR. The state of the universe at a time is given by $(\Psi, Q)$. 
The universal wave function $\Psi$ that evolves unitarily according to the Schr\"odinger equation
\begin{equation}\label{SE}
 i\hbar \frac{\partial \Psi}{\partial t} = H \Psi
\end{equation}
There are actual particles that have precise locations in physical space, represented by $\mathbb{R}^3$. The change in particle configuration $Q = (Q_1, Q_2, ... , Q_N) \in \mathbb{R}^{3N}$ follows the guidance equation (written for the $i$-th particle): 
\begin{equation}\label{GE}
 \frac{dQ_i}{dt} = \frac{\hbar}{m_i} \text{Im} \frac{ \nabla_i \Psi (q) }{  \Psi (q)} (q=Q)
\end{equation}
Moreover, the initial particle distribution is given by the quantum equilibrium distribution: 
\begin{equation}\label{QEH}
\rho_{t_0} (q) = |\Psi(q, t_0)|^2
\end{equation}

 \textbf{W-BM.} Consider an alternative formulation of Bohmian mechanics (W-BM) whose natural interpretation validates DMR.  The state of the universe at a time is given by $(W, Q)$. The universal density matrix $W$ (which can be a mixed state) evolves unitarily according to the Von Neumann Equation: 
\begin{equation}\label{VNM}
i \hbar \frac{\partial W}{\partial t} = [H,  W]
\end{equation}
The particle configuration $Q = (Q_1, Q_2, ... , Q_N) \in \mathbb{R}^{3N}$ follows a new guidance equation \citep{durr2005role}:
\begin{equation}\label{WGE}
\frac{dQ_i}{dt} = \frac{\hbar}{m_i} \text{Im} \frac{\nabla_{q_{i}}  W (q, q', t)}{ W (q, q', t)} (q=q'=Q)
\end{equation}
and is initially distributed by: 
\begin{equation}\label{WBR}
\rho_{t_0} (q) =  W (q, q, t_0) dq
\end{equation}
This version of Bohmian mechanics satisfies equivariance just as the wave-function version does \citep{durr1992quantum, durr2005role}. 

\textbf{$\Psi$-GRW.} There are several versions of GRW theories that validate WFR. In the first one, $\Psi$-GRW0, the fundamental ontology consists only of the universal wave function. The wave function typically obeys (\ref{SE}), but the linear evolution is interrupted randomly (with rate $N\lambda$, where $N$ is the number of particles and $\lambda$ is a new constant of nature of order $10^{-15}$ s$^{-1}$) by collapses: 
\begin{equation}\label{WFcollapse}
\Psi_{T^+} = \frac{\Lambda_{k} (X)^{1/2} \Psi_{T^-} }{||  \Lambda_{k} (X)^{1/2} \Psi_{T^-}  || }
\end{equation}
where $\Psi_{T^-} $ is the pre-collapse wave function,  $\Psi_{T^+} $ is the post-collapse wave function, the collapse center $X$ is chosen randomly with probability distribution $\rho(x) = ||  \Lambda_{k} (x)^{1/2} \Psi_{T^-}  ||^2 dx$,  $k \in \{1, 2, ... N\}$ is chosen  randomly with uniform distribution on that set of particle labels, and the collapse rate operator is defined as:
\begin{equation}\label{collapserate}
\Lambda_{k} (x) = \frac{1}{(2\pi \sigma^2)^{3/2}} e^{-\frac{(Q_k -x)^2}{2\sigma^2}}
\end{equation}
where $Q_k$ is the position operator of ``particle'' $k$, and $\sigma$ is another new constant of nature of order $10^{-7}$ m postulated in current GRW theories. We can also define $\Psi$-GRWf and $\Psi$-GRWm, a flashy version and a mass-density version of GRW with distinct choices of the local beables. For example, $\Psi$-GRWm includes an additional law for the mass-density ontology in spacetime: 
\begin{equation}\label{mxt}
m(x,t) = \bra{\Psi(t)} M(x) \ket{\Psi(t)}
\end{equation}
where $x$ is a physical space variable, $M(x) = \sum_i m_i \delta (Q_i - x)$ is the mass-density operator, which is defined via the position operator $Q_i \psi (q_1, q_2, ... q_n)= q_i \psi (q_1, q_2, ... q_n) $.

\textbf{W-GRW.} Consider a class of alternative formulations of GRW that validate DMR.  For W-GRW0, the fundamental ontology consists only of the universal density matrix $W$, which typically obeys (\ref{VNM}), but the linear evolution is interrupted randomly
\begin{equation}\label{collapse}
W_{T^+} = \frac{\Lambda_{I_{k}} (X)^{1/2} W_{T^-} \Lambda_{I_{k}} (X)^{1/2}}{\text{tr} (W_{T^-} \Lambda_{I_{k}} (X)) }
\end{equation}
with $W_{T^+}$ the post-collapse density matrix, $W_{T^-}$ the pre-collapse density matrix, and  $X$ distributed by the probability density
$\rho(x) = \text{tr} (W_{T^-} \Lambda_{I_{k}} (x))$.

We can similarly define W-GRWf and W-GRWm, the flashy version and the mass-density version \citep{allori2013predictions}. For example,  $W$-GRWm includes the mass-density law: 
\begin{equation}\label{Wmxt}
m(x,t) = \text{tr} (M(x) W(t)).
\end{equation}

\textbf{$\Psi$-EQM.} In the standard version of Everettian quantum mechanics with a fundamental pure state, the universal wave function evolves unitarily by (\ref{SE}), with the branching structure emergent from decoherence. We also have the option to add a separable fundamental ontology in spacetime by defining a mass-density field as in (\ref{mxt}).

\textbf{W-EQM.} In the alternative formulation of Everettian quantum mechanics with a fundamental density matrix, we can unitarily evolve the universal density matrix $W$ by (\ref{VNM}), understand the emergent branching structure via decoherence, and apply self-locating probability to recover the Born rule \citep{ChenChua}.  We also have the option to add a separable fundamental ontology in spacetime by defining a mass-density field as in (\ref{Wmxt}). 

Such theories illustrate the similarities and differences between WFR and DMR. The $\Psi$-theories and the W-theories employ similar equations, but the fundamental objects are different. The $\Psi$-theories postulate a universal wave function, while the W-theories postulate  a universal density matrix that may be a fundamental mixed state. 

For both WFR and DMR, we can postulate other laws to further restrict the nomologically possible quantum states. For example, to explain the manifest arrows of time in $\Psi$ theories,  we may postulate a Past Hypothesis (PH) to constrain the initial wave functions to those compatible with $\Hilbert_{PH}$, a low-dimensional subspace corresponding to the initial low-entropy macrostate  of the universe. W-theories offer some distinctive choices. A particularly natural one is to apply the Initial Projection Hypothesis (IPH) and pick the normalized projection onto $\Hilbert_{PH}$ as the only nomologically possible initial quantum state \citep{chen2018IPH, chen2023detlef}. Such a choice requires DMR (as long as $\Hilbert_{PH}$ has more than one dimension).  

As another example, consider the on-going discussion in quantum cosmology inspired by the No-Boundary proposal \citep{hartle1983wave}. In the WFR framework, we can follow the Hartle and Hawking prescription (insofar as it is well-defined) and select the universal wave function to be the one where space smoothly shrinks to a point towards one temporal boundary. However, there are also proposals that require DMR, such as Page's No-Bang cosmology \citep{page2008no} and the thermal version of the Hartle-Hawking state \citep{barvinsky2006cosmological}. Certain approaches to the black hole information loss paradox require DMR \citep{hawking1976breakdown, hawking1982unpredictability}, but they remain controversial. DMR may also be a natural setting (but not a requirement) for investigating the thermal time hypothesis \citep{connes1994neumann}, according to which the time variable is determined by a universal density matrix. 

\section{Interpretations}

What does a universal density matrix represent in the physical world? The density matrix is an abstract mathematical object. It can be challenging to say what kind of thing it represents.  Nevertheless, the situation is no more difficult than that in the case of WFR. Strategies for interpreting the universal wave function are also available for interpreting the universal density matrix. 

Similar to the situation in WFR \citep{chen2019realism}, we have  four ontological interpretations of the universal density matrix in DMR. 
First, we can understand $W(q,q',t)$ as representing a physical field evolving in a $6N$-dimensional fundamental space represented by $\mathbb{R}^{6N}$. The field assigns properties to every point on that space.  
  Second, we can understand it as representing a low-dimensional multi-field. The fundamental space is a $3$-dimensional space represented by $\mathbb{R}^{3}$, and $W(q,q',t)$  assigns properties to every $2N$-tuple of points on that space. 
  Third, we can understand it as describing properties of spacetime regions. We can obtain, from the universal density matrix, reduced density matrices that correspond to physical properties of  regions in a $4$-dimensional manifold. Such properties are in general non-separable due to quantum entanglement. 
  Finally, we can understand it as representing a geometric object in Hilbert space. 
  
For the ontological interpretations, locality and separability considerations always go together in WFR but may come apart in DMR. For example, in W-BM, the smallest space where everything is separable is $\mathbb{R}^{6N}$, while the smallest space where the dynamics is local is $\mathbb{R}^{3N}$, since the particle configuration lies in $\mathbb{R}^{3N}$. Hence, for those who understand the quantum state as a field-like material object \citep{AlbertEQM, ney2021world}, they face a choice on whether  the fundamental physical space is $3N$-dimensional or $6N$-dimensional. 
  
Alternatively, we can adopt the nomological interpretation of the quantum state. In W-BM, we can regard $W_0$ as a function that prescribes the velocity field for the particle configuration, just like the Hamiltonian function prescribes the velocity field on phase space in classical mechanics. In W-EQM with a mass-density ontology, we have the option of postulating the $m(x,t)$ field as the only material ontology, whose time evolution is given by a law written in terms of a mathematical function corresponding to a universal density matrix.  DMR allows us to choose a simple initial density matrix \citep{chen2018IPH, chen2018HU}. So we have an easy route to the nomological interpretation of the universal quantum state in DMR. 
    

\section{Empirical Equivalence}

In \S3 we saw some examples of physical theories with fundamental universal density matrices that are natural to understand in the DMR framework. Given appropriate choices of physical laws, each density-matrix theory is empirically equivalent to its wave-function counterpart,  so that they cannot be distinguished even in principle by experiment or observation \citep{chen2019quantum1}. 
 
 This is important, for it means the two kinds of theories rise and fall together with respect to empirical evidence. If one is confirmed by evidence, the other is too, and by the same degree. Conversely, if one is disconfirmed by evidence, so is the other. However, the relevant statement of empirical equivalence depends on not just the quantum state but also laws governing its history, and this is a place where different choices for the density-matrix dynamics can make a difference. In \S6, we will consider an empirical difference in the generalized quantum theory  advocated by \cite{cuffaro2021open}. 

There are different ways to understand the empirical equivalence of theories. I shall approach it from three levels. The first  is at the universal level, the second at the subsystem level, and the third at the local level, with significant overlap between the latter two.  Even though wave-function realist theories and density-matrix realist theories are physically distinct, they can be empirically equivalent at all three levels.

\subsection{The Universal Level: Equivalence of Probabilities}

We start with a general conception of empirical equivalence: 

\begin{description}
\item[Criterion for Empirical Equivalence:] Theories $A$ and $B$ are empirically equivalent if they assign the same probability to every possible outcome in every possible measurement.
\end{description}
The criterion should be understood at the universal level, since it concerns all possible measurements that can be done in the universe. If two theories satisfy this criterion, they are empirically equivalent in a very strong sense, concerning not just actual data, but all possible data; not just the actual world, but all possible worlds, permitted by their physical laws. No observation or experiment can distinguish between them. 

The general criterion can take on different forms in different interpretations of quantum mechanics.  For example, we can formulate a special case for Bohmian theories. Since in a Bohmian universe every measurement apparatus is made out of particles with precise positions, every measurement boils down to a position measurement. Hence, to achieve the strong sense of empirical equivalence, we just need two theories to agree on the probability of the particle configurations. 
\begin{description}
\item[Criterion for Empirical Equivalence of Bohmian Theories:] Bohmian theories $A$ and $B$ are empirically equivalent if, for every time $t$, $P_A (Q_t \in dq) = P_B (Q_t \in dq)$.
\end{description}
We can formulate W-BM and $\Psi$-BM  in such a way that the criterion is satisfied. 
\begin{theorem}
Let $W$-BM be the theory of ($W$, $Q$) such that $W$ evolves by (\ref{VNM}), $Q$ evolves by (\ref{WGE}) and satisfies (\ref{WBR}); moreover, a particular  $W(t_0)$ is chosen. Let $\Psi$-BM be the theory of ($\Psi$, $Q$) such that $\Psi$  evolves by the Schr\"odinger equation, $Q$ evolves by the guidance equation and  satisfies the quantum equilibrium distribution in the wave-function version of Bohmian mechanics; moreover, $\Psi(t_0)$ is chosen at random from a statistical ensemble represented by the density matrix $W(t_0)$. $W$-BM and $\Psi$-BM are empirically equivalent in the sense above. 
\end{theorem}

Proof: see \cite{chen2019quantum1}. The basic idea is that, because of equivariance, at every time, the two theories have the same formula for the probability distribution of particle configuration. 

For  Everettian theories, we can draw the same conclusion about the respective theories. Let $W$-EQM be the theory of $W$ such that $W$ evolves by (\ref{VNM}); moreover, a particular  $W(t_0)$ is chosen. Let $\Psi$-EQM be the theory of $\Psi$ such that $\Psi$  evolves by (\ref{SE}); moreover, $\Psi(t_0)$ is chosen at random from a statistical ensemble represented by the density matrix $W(t_0)$. 

$W$-EQM and $\Psi$-EQM are empirically equivalent, because they assign the same probability to every outcome in every experiment. Let $A$ be a self-adjoint operator corresponding to some observable. Suppose  that its spectral measure is given by $\mathscr{A}$, a projection-valued measure. Then the probability that at time $t$,  the outcome of the measurement $x$ will be within some measurable set $M$ is:  
\begin{itemize}
\item  $W$-EQM: $P_{W-EQM} (x \in M) = \text{tr} ( W_t\mathscr{A}(M))$.
\item  $\Psi$-EQM: $P_{\Psi-EQM} (x \in M) = \text{tr} ( W_t\mathscr{A}(M) )$.
\end{itemize}
Under the assumption that probability makes sense in the $\Psi$-Everettian and W-Everettian theories,  the two types of Everettian theories are empirically equivalent. 

For  GRW theories, the $W$ and $\Psi$ versions with different choices of the primitive ontology ($m$ or $f$) are also empirically equivalent to each other (with the arrows denoting empirical equivalence):

\begin{equation}\label{diag}
\xymatrix{
  \Psi\text{-GRWf} \ar[r] \ar[d]   & W\text{-GRWf} \ar[d] \ar[l] \\
  \Psi\text{-GRWm} \ar[r]  \ar[u] & W\text{-GRWm} \ar[u] \ar[l]
}
\end{equation}
For  arguments for why they are empirical equivalent, see \cite{allori2013predictions} and \cite{chen2019quantum1}.


\subsection{The Subsystem Level: Collapses}

The general argument for empirical equivalence is clean and powerful. Nevertheless, it may be useful to analyze how the universe works at the subsystem level and how WFR and DMR can converge on their predictions for the subsystem dynamics. In this section, I shall use Bohmian mechanics as a concrete example, for its sharp mathematical structure offers a particularly clear picture of how subsystem states can differ on $\Psi$-BM and W-BM but lead to the same empirical consequences. 

Following \cite{chen2019quantum1}, consider the following analysis of subsystems in $W$-BM: 

(1) \textbf{Splitting.} For any given subsystem of particles we have a splitting:
\begin{equation}\label{splitting}
q=(x,y),
\end{equation}
with $x$ the generic variable for the configuration of the subsystem and $y$ the generic variable for the configuration of the environment, i.e. the complement of the subsystem. (\ref{splitting}) provides a splitting of the actual configuration into two parts:
\begin{equation}\label{actualsplitting}
Q=(X,Y).
\end{equation}
So we can write the universal density matrix in terms of $W = W(x,y, x',y') $. 

(2) \textbf{Effective density matrix.} 
The subsystem corresponding to the $x$-variables has an \emph{effective density matrix } (at a given time) if the universal density matrix $W(x,y,x',y')$ and the actual configuration $Q=(X,Y)$ (at that time) satisfy: 
\begin{equation}\label{Weffective}
W(x,y,x',y') = \rho(x,x')\gamma(y,y') + W^\bot (x,y, x', y'), 
\end{equation}
such that  
$\gamma(y,y')$ and $ W^\bot (x,y, x', y')$ have macroscopically disjoint $(y,y)$-supports,
and
\begin{equation}\label{supp}
(Y,Y) \in \text{supp } \gamma(y,y'),
\end{equation} 
In this case, the effective density matrix of the subsystem is $\rho(x,x')$. 

(3) \textbf{Conditional density matrix.} The effective density matrix for a subsystem does not always exist. However, we can always define the \emph{conditional density matrix} in the following way:
\begin{equation}\label{conditional}
w(x,x')=W(x,Y,x',Y).
\end{equation}
Here we identify quantum states differing by a constant factor. Given the definition of the velocity (\ref{WGE}), the velocity field of the $x$-system will be given by its conditional density matrix. 

(4) \textbf{Collapse and effective collapse.} When (\ref{Weffective}) and (\ref{supp}) are satisfied, we can  neglect, for all practical purposes, $W^\bot(x,y,x',y')$. The configuration will be carried by the relevant part of the universal density matrix---$\rho(x,x')\gamma(y,y')$---into the future, without much interference from the other parts contained in  $W^\bot(x,y,x',y')$. In this case, we can say that during measurement,  the universal density matrix has undergone an effective collapse from $W_{t^-}$ to $W_{t^+}=\rho(x,x', t^+)\gamma(y,y',t^+)$. Nevertheless, for the subsystem density matrix, represented by the conditional density matrix $w(x,x',t)$, the time evolution is not given by a von Neumann equation but will in measurement-like situation look practically like a collapse, with a rapid evolution to  $w(x,x',t^+)$. 

(5) \textbf{The Fundamental Conditional Probability Formula.} By equivariance, the distribution of $Q_t$ is always given by $W(q,q,t)$. By (\ref{conditional}), at time $t$, for the conditional probability distribution of the configuration of a subsystem $X_t$ given the actual configuration of the environment $Y_t$, we have the fundamental conditional probability formula for $W$ theories: 
\begin{equation}\label{conditionalprob}
P(X_t \in dx | Y_t) = w (x,x,t) dx,
\end{equation}
where $w (x,x',t) = w(x,x',t)^{Y_t} $ is the conditional density matrix of the subsystem at time $t$. Similar to the situation in $\Psi$-BM, the configurations $X_t$ and $Y_t$ are conditionally independent given the density matrix $w (x,x',t)$. 

It is easy to check that pure effective density matrices can emerge following collapses on the conditional states (see \cite{chen2019quantum1} for a simple example). What is surprising is that W-BM offers a new type of collapses that have no analog in textbook quantum mechanics or even $\Psi$-BM. The following example shows in a concrete way how subsystem dynamics can differ on the two theories and yet produce the same probabilistic predictions.\footnote{It was raised as an objection to W-BM by \cite{albert2023guess}, but properly understood it is a feature and not a bug of W-BM.}

Consider two momentum eigenstates, one with momentum $+1$ (moving uniformly to the right) and the other $-1$ (moving uniformly to the left). On $\Psi$-BM, suppose one of them is the actual quantum state guiding the particle. The particle will be either moving to the left or to the right. However, on W-BM, if we regard the equal mixture of the two  as the fundamental density matrix, one might worry that the particle guided by this density matrix will be permanently at rest, because the average of the two gives exactly zero momentum. Nevertheless, when we measure the particle, the record may indicate that the particle is moving.  What is going on? 

 Let us analyze the measurement situation with the subsystem analysis provided above. Let us split the universal configuration $Q$ into $(X,Y)$, with $X$ the subsystem configuration and $Y$ the environmental one. At $t_1$, before measurement, consider two possible universal ``wave functions'' (setting aside the issue of square-integrability): 

\begin{equation}\label{pure9}
 \Psi_A^{t_1}(x, y) = \psi^- (x) \phi^{ready} (y)
\end{equation}
 
\begin{equation}\label{pure10}
 \Psi_B^{t_1}(x, y) = \psi^+ (x) \phi^{ready} (y)
\end{equation}
with $\psi^- (x)$ and $\psi^+ (x)$ denote the momentum eigenstates with momentum $-1$ and momentum $+1$ and $\phi^{ready} (y)$ the quantum state of the environment that is ready to measure the particle's momentum. Let us suppose the two will unitarily evolve into these:
\begin{equation}
 \Psi_A^{t_2}(x, y) = \psi^- (x) \phi^{-} (y)
\end{equation}
 
\begin{equation}
 \Psi_B^{t_2}(x, y) = \psi^+ (x) \phi^{+} (y)
\end{equation}
with $\phi^{-} (y)$ and $\phi^{+} (y)$ the states of the environment indicating the particle in the subsystem is moving to the left and moving to the right, respectively. As usual, we assume that the two records are macroscopically distinct, with $\phi^{-} (y)$ and $\phi^{+} (y)$ having macroscopically-disjoint supports. 

Consider an impure density matrix, the equal mixture of $ \Psi_A^{t_1}(x, y)$ and $ \Psi_B^{t_1}(x, y)$: 
\begin{equation}
 W^{t_1}(x,y,x',y') = \frac{1}{2} \Psi_A^{t_1}(x, y) \Psi_A^{t_1\ast}(x', y') + \frac{1}{2} \Psi_B^{t_1}(x, y) \Psi_B^{t_1\ast}(x', y')
\end{equation}
Let the universal particle configuration be guided by such a density matrix according to the W-BM dynamics. 

Before measurement, at $t_1$, the conditional density matrix of the one-particle subsystem is an impure state: 
\begin{equation}\label{impureExample}
w^{t_1}(x,x') = K W^{t_1} (x,Y^{t_1};x',Y^{t_1})   
=   \frac{1}{2} \psi^- (x) \psi^{-\ast} (x') + \frac{1}{2} \psi^+ (x) \psi^{+\ast} (x')
\end{equation}
with $K$ the normalization constant. 

After measurement, at $t_2$, the subsystem is measured and the record indicates that it is actually moving to the left. The universal quantum state has evolved into this:
\begin{equation}
 W^{t_2}(x,y,x',y') = \frac{1}{2} \Psi_A^{t_2}(x, y) \Psi_A^{t_2\ast}(x', y') + \frac{1}{2} \Psi_B^{t_2}(x, y) \Psi_B^{t_2\ast}(x', y')
\end{equation}
Because of the macroscopically-disjoint supports of $\phi^{-} (y)$ and $\phi^{+} (y)$, the two terms on the right hand side, $\Psi_A^{t_2}(x, y) \Psi_A^{t_2\ast}(x', y')$ and $\Psi_B^{t_2}(x, y) \Psi_B^{t_2\ast}(x', y')$, also have macroscopically-disjoint supports. 

At $t_2$, the environmental configuration $Y^{t_2}$ lies within the support of $\phi^-(y)$. Plugging it into the universal density matrix, we have (almost) zero contributions from terms with $\phi^+(y)$. So the conditional density matrix of the subsystem becomes a pure state: 
\begin{equation}
w^{t_2}(x,x') = K' W^{t_2} (x,Y^{t_2};x',Y^{t_2})   
=   \psi^- (x)\psi^{-\ast} (x') 
\end{equation}
with $K'$ the normalization constant. 
The evolution of the conditional quantum state of the subsystem from $w^{t_1}(x,x')$ into $w^{t_2}(x,x')$ is a  non-unitary one, corresponding to a more general type of quantum state collapses. The collapse into $\psi^- (x)\psi^{-\ast} (x') $ has a probability of $1/2$, the same as the collapse into  $\psi^+ (x)\psi^{+\ast} (x') $. Notice that the collapse  \textit{takes a mixed state to a pure state}. It is a real collapse for the subsystem state in W-BM. However, this is a case where textbook QM and $\Psi$-BM do not postulate a collapse at all; they would regard the transition as an epistemic change of learning which of the two pure states is the actual one, or, in Bayesian terms, update by conditionalization.  Let us call this \textit{collapse without ``textbook collapse.''} It illustrates a crucial difference between $\Psi$-BM and W-BM at the subsystem level. In this case, the conditional state collapses in W-BM but the corresponding one does not collapse in $\Psi$-BM, and yet they are predictively equivalent, assigning the same probabilities to possible outcomes. In other words, they offer different physical explanations for the empirical phenomena.


\subsection{The Local Level: Canonical Typicality}

The previous arguments are quite general and sufficient to establish exact empirical equivalence with exact agreements of probabilities.  I shall propose another argument for an approximate sort of empirical equivalence between WFR and DMR based on local indistinguishability of ``small'' subsystems in a ``large'' universe. It exploits recent results in quantum statistical mechanics known as \textit{canonical typicality}. Its scope is more restricted:  the argument depends on the size of the actual universe and how much of it we have direct empirical access. However, because of the generous bound on subsystem size, it presumably applies to our actual epistemic situation. I include the argument here because it is surprising and interesting. 

In the early 2000s, several teams of researchers \citep{gemmer2003distribution, goldstein2006canonical, popescu2006entanglement}  have independently discovered a feature of quantum mechanics that has no parallel in classical mechanics. Suppose we have a large quantum universe $U$ with a wave function $\Psi$, with the universe partitioned into a small subsystem $S$ and its environment $E$ (the complement of $S$ in $U$) and the universal Hilbert space is the tensor product of the Hilbert spaces of $S$ and $E$, i.e. $\Hilbert = \Hilbert_S \otimes \Hilbert_E$. Suppose further that there is a global constraint $R$ on the universe such that  $\Psi \in \Hilbert_R$, with $\Hilbert_R$ the subspace of the universal Hilbert space corresponding to $R$. The reduced density matrix of the subsystem $S$, obtained from $\Psi$ by tracing out the environmental degrees of freedom in $E$, is
\begin{equation}
  \rho^{\Psi}_S= \tr_E |\Psi\rangle\langle\Psi| 
\end{equation}
Now, the normalized projection onto $\Hilbert_R$ is $W_R = \mathbb{I}_R / \dim\Hilbert_R$, with $\mathbb{I}_R$ the projection  onto $\Hilbert_R$. $W_R$ is mathematically the same as the statistical density matrix  corresponding to the uniform probability distribution, with respect to the normalized surface area measure $\mu$, on the unit sphere of $\Hilbert_R$:
\begin{equation}
  W_R = \int_{\mathbb{S}(\Hilbert)}\mu(d\psi) \, |\psi\rangle\langle\psi|\
\end{equation}
The reduced density matrix of the subsystem $S$, obtained from $W_R$ by tracing out the environmental degrees of freedom in $E$, is
\begin{equation}
  \rho^{W_R}_S= \tr_E W_R
\end{equation}
Remarkably, it is shown that, under suitable conditions and for typical wave functions in $\Hilbert_R$,
\begin{equation}
    \rho^{\Psi}_S \approx   \rho^{W_R}_S
\end{equation}
where the approximation can be made precise by various inequalities \citep{teufel2024canonical}. This is known as canonical typicality. The phenomenon is a manifestation of quantum entanglement and has no classical analogue. 

\cite{popescu2006entanglement} gloss canonical typicality as ``almost every pure state of the universe is locally (that is, on the system) indistinguishable from [$W_R$].'' They interpret the latter as encoding a statistical density matrix. Results of this form have been used to support the individualistic picture of quantum statistical mechanics that it is meaningful to use standard statistical ensembles for subsystem descriptions even when the universe is in a pure state, and to justify the postulate of micro-canonical ensembles and canonical ensembles.  

However, there is an under-appreciated corollary of canonical typicality, relevant to the empirical equivalence between WFR and DMR.  Precisely because of canonical typicality, we cannot use local data to distinguish whether the universal quantum state is pure or impure. 

How close  $\rho^{\Psi}_S $  and $ \rho^{W_R}_S $ are to each other depends  on the size of the subsystem and the size of the environment. To apply canonical typicality, the subsystem can contain up to 50\% of the number of degrees of freedom of the total system (e.g. the universe) \citep{teufel2024canonical}. Let’s consider the solar system as such a ``small'' subsystem of the universe. The spatial region of the solar system includes all our direct evidence of the universe collected so far. We may have other observational data about the universe, such as other galaxies, but it will be inferred from our direct evidence (pointer readings), which is locally here in the solar system. Given the local data in the solar system, we only have access to the reduced density matrix of the solar system. The latter, by canonical typicality, is more or less the same whether it is obtained from a typical wave function in the present macrostate of the universe $\Hilbert_M$ or the normalized projection onto the subspace associated with the macrostate, a mixed state. Therefore, given the local data we have, we cannot tell whether the universe is in a pure state or a mixed state.\footnote{It is true that we often know more about the subsystem state than that it is reduced from a typical pure state of the present macrostate.  The argument can be modified to accommodate such information.  First, restrict ourselves from the present macrostate $\Hilbert_M$ to those wave functions whose earlier histories are compatible with a Past Hypothesis $\Hilbert_{PH}$. This gives us a smaller subspace $\Hilbert_{M'}$. Next, project the wave functions onto the subspace corresponding to the present ``branch'' of the wave function, say, with humans existing. This yields another subspace $\Hilbert_{M''}$. How the projection is physically implemented depends on the interpretation of quantum mechanics. (In Bohmian theories, we can obtain a much more detailed conditional density matrix of the subsystem, using the environmental configuration located outside the subsystem of interest.) Once we focus on $\Hilbert_{M''}$, the argument from canonical typicality applies again. }

We are  gradually exploring other parts of the universe. But it is realistic to expect that we will not have direct empirical access to more than 50\% of the microscopic degrees of freedom in the universe.  Considered as an abstract subsystem, our evidence by itself will (typically) not be able to distinguish between a universal pure state or a universal mixed state. In other words, the totality of direct empirical evidence, as long as it is sufficiently local, will typically (according to the standard measure) underdetermine the universal quantum state. This argument  requires that our evidence be local in the appropriate sense. As it turns out, by appealing to another result called \textit{distribution typicality}, we can drop the locality assumption and generalize the argument to any empirical evidence \citep{chen2024canonical}. 


\section{Generalizations}

 In \S3 and \S5, I have considered density-matrix realist theories that are ``conservative extensions'' from their wave-function realist counterparts. The laws in the former are natural generalization of those in the latter. For example, (\ref{VNM}) conservatively extends from  (\ref{SE}),  (\ref{WGE})  from (\ref{GE}), and (\ref{collapse}) from  (\ref{WFcollapse}). All of those laws may be called closed-system dynamics. They differ from the more general open-system dynamics suggested by \cite{cuffaro2021open}.  Regarding my definition of DMR in \S2, the relevant quantum state whose objectivity is at issue is the universal one. With the focus on closed-system dynamics and universal quantum states, we can establish the empirical equivalence between WFR and DMR. For clarity, let me label the version of DMR I am committed to as DMR$^U_C$, with $U$ standing for the universe and $C$  closed-system dynamics. 

In this section, let us consider two generalizations of DMR. They are related to the proposals of \cite{robertson2022search} and \cite{cuffaro2021open}. There are important similarities and differences. 

\subsection{The Subsystem Version: DMR$^S$}

First, consider DMR$^S$, the thesis that the reduced density matrices assigned to subsystems of the universe are objective. This view faces an apparent objection. Since reduced states are obtained by tracing out the environment, which is akin to taking a statistical average, they surely cannot be entirely objective. The standard view regards the reduced states as carrying incomplete information about subsystems.  Against the standard view, \cite{robertson2022search}  argues for  DMR$^S$ by suggesting that density matrices are inevitable and the most complete description for subsystems entangled with their environments, concluding that density matrices should not be understood as entities that encode subjective information but as states of the relevant physical systems\footnote{For a similar argument, see \cite{wallace2016probability}.}: 
\begin{quotation}
  The density matrix is arguably the best mathematical object to represent the state of the individual system (rather than the wavefunction $\psi$), since $\rho$ is a more general object than $\psi$. Quantum systems rapidly become entangled with their environment — which means that the individual system cannot be described by a wavefunction, but instead must be a (reduced) density matrix (by tracing over the environment). Since the density matrix formalism is more general, and sometimes required, the density matrix should be taken to be more fundamental...... Thus the individual state of the system in QM is not represented by a ray in Hilbert space (the quantum equivalent of a point in phase space), but a density matrix.... Thus, there is no difference in the mathematical object that represents the state of the individual system, and a probability distribution over it. Thus, in QSM, the dichotomy between ‘being a property of a probability distribution’ and ‘being a property of the individual system’ never arises.
\end{quotation}
But the defender of the standard view has a ready response: the dichotomy can still arise in quantum statistical mechanics. For example, even though the hydrogen atom is entangled with the rest of the universe,  we attribute to it a pure state, while the reduced density matrix is heavily mixed.  How different quantum theories explain this phenomenon is briefly discussed in \S5.3. If WFR is correct, in Bohmian mechanics the subsystem can possess a (conditional) pure state, obtained from plugging the environmental configuration into the universal wave function. Even on DMR$^U_C$, such a conditional state obtained from the universal density matrix can still be pure. (The Everettian story will involve branching.) 

Hence, there is still a choice. Empirically equivalent but physically distinct theories can assign different subsystem states, some pure and some mixed. One has the option to regard the mixedness as a symptom for non-objectivity. Hence, we are not forced to accept DMR$^S$. 

However, if we accept DMR$^U_C$, there is a natural argument for a restricted version of DMR$^S$.  If the universal quantum state is mixed, the actual states of \textit{some} subsystems will be more mixed than when we assume WFR.   Equation (\ref{impureExample}) in \S5.2 provides an example. See \cite[sect.4]{durr2005role} for another type of example involving spin. Such mixed states can be objective, though not as fundamental as the universal density matrix. Hence, we may have motivations to adopt the restricted version of DMR$^S$, treating some subsystem density matrices as objective (though not  fundamental), if we already accept DMR$^U_C$. 

\subsection{The Open Systems View: DMR$^U_O$}

Next, consider DMR$^U_O$, the idea that the universal density matrix is objective, can be impure, and evolves non-unitarily according to a fundamental open-system dynamical equation, such as the Lindblad equation \citep{cuffaro2021open}.  For example, one can formulate a Bohm-type theory where particles exist in 3-dimensional space and are guided by a universal density matrix that evolves according to not the von Neumann equation (\ref{VNM}) but a Lindblad equation. The predictions of the theory differs from those of $\Psi$-BM and $W$-BM.\footnote{This theory is suggested but not endorsed in \cite[sect.4.3]{allori2013predictions}. Thanks to Roderich Tumulka for discussions about this point.}  Advocates of the open-system dynamics should be understood as committing not just to the theoretical possibility of such open-system dynamics, but also assigning a non-zero epistemic probability that it is the correct description of the fundamental law of the universe. Otherwise, it will be epistemically equivalent to holding DMR$^U_C$. 

Thus understood, DMR$^U_O$ is distinct from DMR$^U_C$.\footnote{DMR$^U_C$ includes W-GRW theories, but their fundamental laws differ from the master equations of open systems, such as the Lindblad equation. See \cite[sect. 6.3]{goldstein2012quantum}.} While DMR$^U_C$ can be shown to be empirically equivalent as WFR, because the fundamental dynamics can be of the same type, advocates of DMR$^U_O$ should assign a non-zero probability in predictions that diverge from standard formulations of realist quantum theories. This provides an epistemic argument in favor of DMR$^U_C$ and against DMR$^U_O$. Insofar as all data are in full agreement with quantum mechanics, DMR$^U_O$ makes the data more surprising than DMR$^U_C$. Schematically, we may represent the situation as follows: 

\begin{equation}
  \Prob(E_{QM} \, | \, \text{DMR}^U_C) >   \Prob(E_{QM} \, | \, \text{DMR}^U_O)
\end{equation}
and it is rationally permissible to have the following prior assignment: 
\begin{equation}
  \Prob( \text{DMR}^U_C) =   \Prob(\text{DMR}^U_O)
\end{equation}
We are also allowed to assign higher prior in  DMR$^U_C$ than  DMR$^U_O$. The reason we are not rationally required to assign equal or higher prior in the latter, even though every model of  DMR$^U_C$ will correspond to a model of  DMR$^U_O$ but not vice versa, is because they are distinct hypotheses about the fundamental laws \citep{chen2023simplicity, chenCUP}. Hence, by Bayes's rule, the posterior probability in DMR$^U_C$, given available data, is higher than that in DMR$^U_O$. 

In other words, DMR$^U_C$ is the conservative and empirically equivalent extension of WFR, so that they are confirmed and disconfirmed to the same degree by empirical evidence, while DMR$^U_O$ takes on extra epistemic risks. If empirical evidence had been different and if we have seen deviations from quantum mechanics, we might have possessed empirical evidence to favor DMR$^U_O$ over DMR$^U_C$. But insofar as empirical evidence is compatible with DMR$^U_C$ and WFR and incompatible with certain versions of DMR$^U_O$, the opposite is the case. 

This does not mean it is not  pragmatically fruitful to consider  DMR$^U_O$. Clarifying such a framework can be useful for developing possible quantum gravity or post-quantum theories. But insofar as no empirical evidence deviates from quantum predictions, we do not have an epistemic reason to go beyond DMR$^U_C$.

\section{Frequently Asked Questions}

In this section, I answer six sets of frequently asked questions regarding DMR, by pointing to relevant discussions in earlier sections and elsewhere in the literature. 

\subsection{Intelligibility}

How is DMR an intelligible thesis? Density matrices, by definition, are about partial information or epistemic ignorance. It makes no sense to be a realist about the density matrix. 

Answer: Density matrices can play different roles in quantum theory. While some of them encode partial information or epistemic ignorance, it does not follow that all density matrices play such roles. The universal quantum state can be a fundamental mixed state, not arising from entanglement or statistical mixture, as shown in \S3-4. 

\subsection{Familiarity}

DMR is unfamiliar. No textbooks in quantum mechanics introduce density matrices as fundamental features of the universe. Why should we consider DMR? 

Answer: The lack of familiarity of DMR is not an epistemic reason for taking it less seriously.  We can come to a better understanding of DMR by thinking through the definition, examples, interpretations, possible generalizations, and relation to WFR. Insofar as we do not have decisive reasons against DMR, we should consider DMR as a serious alternative to WFR. In \S7.6, I also mention some reasons to favor DMR over WFR. 

\subsection{Purification}

Even if DMR is intelligible and there may be reasons to consider it, we can always purify the mixed state by adding a factor in a larger Hilbert space. Why can't we just focus on WFR? 

Answer: From the perspective of WFR, that is indeed possible. However, absent any reason to think WFR has to be true, DMR is on the table. From the perspective of DMR,  the purification is not always physical. The universal state may just be a fundamental mixed state that does not arise from entanglement from the outside, as there is no physical degree of freedom that it is entangled with.  

\subsection{Physical Equivalence}

If DMR and WFR are empirically equivalent, are they not the same theory? How are they physically or ontologically distinct? 

Answer: They are physically distinct theories,  as they prescribe different physical possibilities. Are fundamental universal mixed states physically possible?  Yes on DMR but No on WFR.  Empirical equivalence does not entail physical or ontological equivalence. Even though DMR and WFR can yield the same predictions, they provide different explanations for how such phenomena arise. For example, a universal mixed state can give rise to particle trajectories, quantum state collapses, and multiverse evolutions that are ontologically different from those arising from a universal pure state.  

\subsection{Occam's Razor}

Doesn't Occam's Razor favor WFR over DMR? Isn't DMR more complicated than WFR? 

Answer: The wave-function realist theories and density-matrix realist theories discussed in \S3 employ very similar laws. For example, (\ref{VNM}),  (\ref{WGE}),  (\ref{collapse}) are no more complex than their wave-function counterparts (\ref{SE}),  (\ref{GE}), and  (\ref{WFcollapse}). Regarding state space structure, both wave functions and density matrices live in the Hilbert space. One may insist that, intuitively speaking, there are more density matrices than wave functions, since every pure state corresponds to a one-dimensional projection density matrix but not every mixed state has a corresponding wave function. Hence, one might make a modal version of Occam's Razor by objecting that WFR is more modally parsimonious. The modal Razor might work when considering WFR and DMR without any additional laws. However, the argument can work against WFR, because some versions of DMR are more modally parsimonious than versions of WFR. For example, the Wentaculus (a version of DMR) is compatible with exactly one initial quantum state but the quantum Mentaculus (a version of WFR) is compatible with infinitely many. See \cite{chen2023detlef}.  

\subsection{Theoretical Payoffs}

What are the theoretical payoffs of DMR? 

Answer: We should take DMR at least as seriously as WFR, regardless of the theoretical payoffs of the former. Since DMR is a clear and intelligible alternative to WFR, and since it is no more complicated than WFR, both of them should be on the table. That being the case,  there are many theoretical payoffs of DMR. I find the following most compelling, but it is by no means an exhaustive list:
\begin{itemize}
  \item \textit{Nature of the quantum state.} DMR provides a new argument for the nomological interpretation of the quantum state.  On the nomological interpretation, the universal quantum state is regarded as a nomic object like the Hamiltonian telling objects how to move. However, the exact initial wave function may not be sufficiently simple to be a law.  DMR offers a new possibility \cite{chen2018IPH, chen2018HU, chen2023detlef}. For example, in the Wentaculus theories, we can stipulate that the initial quantum state is the normalized projection onto the Past Hypothesis subspace, one that is no more complex than the Past Hypothesis itself. Insofar as PH can be simply specified, the initial density matrix can be too.
  \item \textit{Strong Determinism.} DMR can potentially satisfy our explanatory desire to have all physical facts explained by simple laws. Determinism provides conditional explanations of later states in terms of earlier ones. But that does not explain the initial state of the universe, which requires an additional postulate. With Everettian Wentaculus, the initial density matrix is pinned down by a law, and the multiverse evolution is too \citep{chen2022strong, ChenNature2023}. 
  \item \textit{Statistical Mechanical Probability.} Insofar as we have a fundamental density matrix in DMR that can mimic the statistical density matrix over the underlying pure state of WFR, we can regard the statistical mechanical probability distribution as being subsumed under the fundamental density matrix and become part of the quantum mechanical probability \citep{chen2018valia}.  
  \item \textit{Theoretical Unity.} In a universe with a fundamental wave function, most subsystems may be in mixed states but the universal one is pure. This is not the case in a universe with a fundamental mixed-state density matrix, where typical subsystems and the universe are all in mixed states \citep{chen2018IPH}. 
  \item \textit{Nomic Exactness.} DMR provides a new route to eliminate fundamental nomic vagueness, i.e. vagueness in the fundamental laws of nature \citep{chen2018NV}. 
\end{itemize}

\section{Conclusion}

 DMR is a clear and intelligible thesis. With suitable choices of states and dynamics, DMR is empirically equivalent to WFR.  Since they are compatible with different sets of states, they cannot both be correct. It is therefore an interesting case study for scientific realism.  In a quantum universe, we face a substantive and interesting choice regarding what kind of quantum theory we should accept, even after choosing our favorite interpretation of quantum theory. Empirical evidence underdetermines between $\Psi$-theories and W-theories, so our decision needs to rely on theoretical virtues (super-empirical considerations).  In a quantum universe, a satisfactory defense of scientific realism must confront this issue and justify what reasons (if any) we have to favor one over the other. 

\section*{Acknowledgement}

For helpful discussions, I am grateful to David Albert, Craig Callender, Eugene Chua, Mike Cuffaro, Sheldon Goldstein, Katie Robertson, Stephan Hartmann, and Roderich Tumulka. I thank the editors and two anonymous reviewers for valuable feedback. This project is supported by an Academic Senate Grant from University of California, San Diego and Grant 62210 from the John Templeton Foundation. The opinions expressed here are those of the author and do not necessarily reflect the views of the John Templeton Foundation.


\bibliography{test}


\end{document}